# Dark-field signal extraction in propagation-based phase-contrast imaging

T E Gureyev[1,2,3,4], D M Paganin[3], B Arhatari[1,5,6], S T Taba[2], S Lewis[2], P C Brennan[2] and H M Quiney[1,2]

[1] School of Physics, the University of Melbourne, Parkville 3010, Australia
[2] Faculty of Health Sciences, the University of Sydney, Lidcombe 2141, Australia
[3] School of Physics and Astronomy, Monash University, Clayton 3800, Australia
[4] School of Science and Technology, University of New England, Armidale 2351, Australia
[5] Australian Synchrotron, ANSTO, Clayton 3168, Australia
[6] Department of Chemistry and Physics, La Trobe University, Bundoora 3086, Australia

**Abstract**

A method for extracting the dark-field signal in propagation-based phase-contrast imaging is proposed. In the case of objects consisting predominantly of a single material, or several different materials with similar ratios of the real decrement to the imaginary part of the complex refractive index, the proposed method requires a single image for extraction of the dark-field signal in two-dimensional projection imaging. In the case of three-dimensional tomographic imaging, the method needs only one image to be collected at each projection angle. A preliminary example demonstrates that this method can improve the visualization of microcalcifications in propagation-based X-ray breast cancer imaging. It is suggested that the proposed approach may be useful in other forms of biomedical imaging, where it can help one to obtain additional small-angle scattering information without increasing the radiation dose to the sample.

**1. Introduction**

Hard X-ray phase-contrast imaging has become a popular method for non-destructive imaging of samples consisting of materials with small differences in X-ray absorption properties - see, for example, the original papers (Goetz *et al* 1979, Ingal *et al* 1995, Davis *et al* 1995, Momose 1995, Snigirev et al 1995, Momose *et al* 1996, Wilkins *et al* 1996, Nugent *et al* 1996, David *et al* 2002, Momose *et al* 2003, Pfeiffer *et al* 2008, Olivo *et al* 2011, Munro *et al* 2012, Wen *et* al 2010, Morgan *et al* 2011a, Bérujon *et al* 2012, Morgan *et al* 2012) or more recent reviews (Wilkins *et al* 2014, Wen 2019). Importantly, while hard X-rays with wavelength of 1 Å or shorter are typically needed in order to image the internal structure of biomedical samples with thickness of several centimeters, the resultant two-dimensional (2D), e.g. projection, images, or three-dimensional (3D), e.g. Computed Tomography (CT), images, tend to display weak contrast between different types of soft tissue since they tend to have similar X-ray absorption properties. In this situation, phase-contrast imaging which utilizes refraction, as well as absorption, of X-rays in the samples, has been shown to significantly increase the soft-tissue contrast under suitable experimental conditions. Since biomedical samples are often radiation sensitive, the X-ray dose also needs to be carefully controlled, and the corresponding "figure of merit" is often represented by the contrast-to-noise ratio (CNR) or signal-to-noise ratio (SNR) per unit dose (Diemoz *et al* 2012, Wilkins *et al* 2014). However, this "figure of merit" still does not present a complete picture, since the CNR and SNR can always be increased at the expense of spatial resolution, without changing the dose. Therefore, in order to evaluate the effectiveness of a particular X-ray imaging method or apparatus, it is usually necessary to take into account all three of these key parameters: CNR/SNR, X-ray dose and spatial resolution. One imaging quality characteristic that takes into account all these factors simultaneously and is invariant with respect to simple operations such as binning of the detector pixels (which increases the SNR, but lowers the spatial resolution), is represented by "intrinsic imaging quality" (Gureyev et al 2014, Gureyev et al 2019). Other image quality metrics can be also used in this context (Wilkins et al 2014, Mait et al 2018).

Different modalities of X-ray phase-contrast imaging have been studied extensively in recent years, both theoretically and experimentally. The simplest such modality is propagation-based imaging (PBI) (Snigirev et al 1995, Wilkins et al 1996, Nugent et al 1996), which utilizes free-space propagation of spatially coherent transmitted X-rays from the sample to the detector as a means for visualizing the phase contrast, i.e. transforming local variations of the phase shifts in the transmitted X-ray beam in the "object" plane into detectable intensity variations in the "image" plane downstream the optical axis of the imaging system. Analyzer-based imaging (ABI) (Goetz et al 1979, Ingal et al 1995, Davis et al 1995) uses Bragg or Laue diffraction of the transmitted beam in a high-quality "analyzer" crystal in order to render the phase contrast detectable. An even more experimentally demanding, but potentially more sensitive, method is based on the use of Bonse-Hart type X-ray crystal interferometers (Momose 1995, Momose et al 1996). Other approaches, which are currently being actively developed for biomedical phase-contrast X-ray imaging include the grating-interferometer-based (GIB) techniques (David et al 2002, Momose et al 2003, Pfeiffer et al 2008) and various forms of coded-aperture type methods, including the "edge illumination" (EI) method (Olivo et al 2011, Munro et al 2012), two-dimensional gratings (GB) (Wen et al 2010, Morgan et al 2011a, Morgan et al 2011b) and speckle-based (SB) methods (Bérujon et al 2012, Morgan et al 2012).

Most of the X-ray phase-contrast imaging methods listed above, with the notable exception of PBI, have been used both in the bright-field and dark-field modes, i.e. respectively with or without the zero-spatial-order transmitted beam contributing to the registered image (Ingal et al 1995, Pfeiffer et al 2008, Munro et al 2012, Wilkins et al 2014). The dark-field mode tends to naturally emphasize the small-angle scattering (SAXS) signal, in particular (Davis 1994, Pfeiffer et al 2008, Wilkins et al 2014, Arfelli et al 2018). The methods which use gratings, coded apertures or analyzer crystals can select between the bright-field and the dark-field modes easily and naturally by changing the effective angular transmission bandwidth of the "wavefront analyzer" element with respect to the propagation direction of the undiffracted beam (zero-order Fourier component). A similar selection cannot be achieved in the PBI method since it does not involve any optical elements and therefore always includes all diffraction orders, up to the highest order determined by the resolution limit of the imaging setup, in the registered image. A very recent study however has demonstrated, via a Fokker-Planck approach (Morgan et al 2019, Paganin et al 2019), that the extraction of SAXS signal in PBI may be possible by means of numerical post-processing of two or more images collected at different propagation distances under suitable conditions. In the present work we demonstrate that it may be possible to extract the dark-field signal, containing the SAXS contribution, using a single PBI image, provided that 1) the spatial resolution of the imaging system is sufficiently high, and 2) the sample may be considered "homogeneous" in the sense that the ratio of the real decrement to the imaginary part of the complex refractive index is the same at any point inside the object (Paganin et al 2002). The class of homogeneous objects (sometimes also called "monomorphous") includes all objects consisting of a single material (the density of these objects is still allowed to vary throughout the volume), and all objects consisting of low-Z elements (Z < 10) when illuminated by X-rays with energy approximately above 60 keV (Wu et al 2005). It has been also demonstrated that many biomedical samples consisting of different soft tissues (e.g. healthy and malignant breast tissues) may be considered approximately homogeneous for the purpose of quantitative analysis of PBI images (Gureyev et al 2013).

## 2. Dark-field signal in PBI

Let an object (scatterer) be located in the vicinity of the optic axis in the half-space $z < 0$ immediately before the 'object' plane $z = 0$. We assume for simplicity that the wave incident on the sample is a plane monochromatic wave with wavelength $\lambda$ and unit intensity, propagating along the optic axis $z$, i.e. the complex amplitude of the incident wave is $\exp(ikz)$, $k = 2\pi/\lambda$. Generalization of the following results to cases involving polychromatic and spatially partially coherent incident radiation can be carried out similarly to the way described in (Gureyev et al 2006). The scattering properties of the object are assumed to be such that the wave transmitted through the object is paraxial, i.e. all the wavefront normals in the object plane are contained in a narrow cone around the direction of the $z$ axis. The transmitted wave propagates in the free half-space $z > 0$ until it reaches a position-sensitive detector. As the transmitted wave has been assumed to be paraxial, its evolution in the free half-space $z > 0$ can be described by the Fresnel integral (Paganin 2006),

$$U_R(x,y) = \frac{\exp(ikR)}{i\lambda R} \iint \exp\{\frac{i\pi}{\lambda R}[(x-x')^2 + (y-y')^2]\} U_0(x',y') \, \mathrm{d}x' \mathrm{d}y', \qquad (1)$$

where $U_0(x,y) \equiv a(x,y)\exp[i\varphi(x,y)]$ is the transmitted complex scalar amplitude of the wave in the object plane having transverse coordinates $(x, y)$ and $R$ is the distance between the object and image planes. The detector is assumed to be capable of measuring the spatial distribution of intensity in the image plane,

$$I_R(x,y) = |U_R(x,y)|^2. \qquad (2)$$

In phase-contrast imaging and phase-contrast tomography one is often interested in finding the object-plane phase $\varphi(x,y)$ and intensity $I_0(x,y) = a^2(x,y)$ from the measured intensity distribution in one or more image planes (Paganin 2006). It is easy to see that Eq. (2) is non-linear with respect to the object-plane phase and amplitude, and as such may be challenging to solve analytically or numerically. For example, the well-known family of Gerchberg-Saxton type phase retrieval algorithms (Fienup 1987) does not work well in the near-Fresnel region (Gureyev 2003). On the other hand the linearized form (approximation) of Eq. (2) that will be employed below, which can be sufficiently accurate under certain well-specified conditions, can also be convenient for use in the retrieval of object-plane phase and amplitude in the Fresnel region (Gureyev et al 2004).

Let us assume that it is possible to represent the complex wave amplitude in the object plane in the following form:

$$U_0(x,y) = U_{TIE}(x,y) U_{Born}(x,y), \qquad (3)$$

where $U_{TIE}(x,y) \equiv \exp[-\mu_{TIE}(x,y) + i\varphi_{TIE}(x,y)]$ is a slowly varying "envelope" function satisfying the validity conditions for the Transport of Intensity equation (TIE) (Teague 1983), and $U_{Born}(x,y) = \exp[-\mu_{Born}(x,y) + i\varphi_{Born}(x,y)]$ contains small but rapidly varying absorption and phase shift components, such that $|\mu_{Born}| \ll 1$ and $|\varphi_{Born}| \ll 1$, and hence $U_{Born}(x,y) \cong 1 - \mu_{Born}(x,y) + i\varphi_{Born}(x,y)$. The term $U_{Born}(x,y)$ may contain the SAXS signal, in particular. Note that the "Born" subscript here refers to the first Born approximation, which is applicable in weak-scatter contexts (Paganin 2006). The following normalization can be also always imposed: $\iint I_{TIE}(x,y)\mu_{Born}(x,y)dxdy = \iint I_{TIE}(x,y)\varphi_{Born}(x,y)dxdy = 0$. Under these conditions it is possible to show (Gureyev et al 2004) that

$$\hat{I}_R(\xi,\eta) = \hat{I}_{TIE}(\xi,\eta) - (R/k)[\nabla_\perp \cdot (I_{TIE}\nabla_\perp \varphi_{TIE})]\hat{\,}(\xi,\eta) - \\ 2\cos[\pi\lambda R(\xi^2+\eta^2)](I_{TIE}\mu_{Born})\hat{\,}(\xi,\eta) + 2\sin[\pi\lambda R(\xi^2+\eta^2)](I_{TIE}\varphi_{Born})\hat{\,}(\xi,\eta), \qquad (4)$$

where the overhead hat symbol denotes the two-dimensional Fourier transform with respect to $x$ and $y$, $\hat{f}(\xi,\eta) = \iint \exp[-i2\pi(x\xi+y\eta)]f(x,y)dxdy$. The first two terms on the right-hand side of eq.(4) correspond to the homogeneous TIE (TIE-Hom) approximation (Paganin et al 2002) of the intensity distribution in the image plane $z = R$:

$$\hat{I}_{R,TIE}(\xi,\eta) \equiv \hat{I}_{TIE}(\xi,\eta) - (R/k)[\nabla_\perp \cdot (I_{TIE}\nabla_\perp \varphi_{TIE})]\hat{\,}(\xi,\eta). \qquad (5)$$

Since $(I_{TIE}\mu_{Born})\hat{\,}(0,0) = (I_{TIE}\varphi_{Born})\hat{\,}(0,0) = 0$ due the chosen normalization, the last two terms of eq.(4) correspond to the dark-field part of the first Born approximation for the following intensity distribution in the image plane:

$$\hat{I}_{R,Born}(\xi,\eta) \equiv I_0\delta(\xi,\eta) + \hat{I}_R(\xi,\eta) - \hat{I}_{R,TIE}(\xi,\eta) = I_0\delta(\xi,\eta) - \\ 2\cos[\pi\lambda R(\xi^2+\eta^2)](I_{TIE}\mu_{Born})\hat{\,}(\xi,\eta) + 2\sin[\pi\lambda R(\xi^2+\eta^2)](I_{TIE}\varphi_{Born})\hat{\,}(\xi,\eta), \qquad (6)$$

where $I_0 \equiv \iint I_0(x,y)dxdy$ (although this particular choice of constant $I_0$ in eq.(6) is not essential).

Equations (4)-(6) suggest the following phase retrieval procedure (Gureyev 2004), which is reminiscent of the well-known Gerchberg-Saxton-Fienup type phase retrieval algorithms (Gerchberg et al 1972, Fienup 1987). First, we take an initial approximation $\hat{I}_R(\xi,\eta) \approx \hat{I}_{R,TIE}(\xi,\eta)$ and solve the TIE eq.(5) for $I_{TIE}$ and $\varphi_{TIE}$. In the second step, we take the difference $\hat{I}_R(\xi,\eta) - \hat{I}_{R,TIE}(\xi,\eta)$ and solve the conventional first Born (also known as "Fourier optics") equation, i.e. eq.(6), for $I_{TIE}\mu_{Born}$ and $I_{TIE}\varphi_{Born}$. The complex amplitude in the object plane is then obtained from eq.(3). The second step can be easily iterated multiple times, if desired (Gureyev et al 2004). This algorithm becomes simpler in the case of homogeneous objects, where it requires only a single defocused image for the reconstruction of both the absorption and the phase shift in the object plane.

As mentioned in the Introduction, homogeneous objects are characterized by the proportionality relationship between the attenuation and phase in the object plane, $\varphi_0(x,y) = (\gamma/2)\ln I_0(x,y)$, with the proportionality constant $\gamma$ equal to the ratio of the real decrement to the imaginary part of the complex refractive index $n(x,y,z) = 1 - \delta(x,y,z) + i\beta(x,y,z)$ inside the object, i.e. $\gamma = \delta(x,y,z)/\beta(x,y,z) = const$ (Paganin et al 2002, Paganin et al 2004). In the case of a homogeneous object, eq.(4) becomes

$$\hat{I}_R(\xi,\eta) = \hat{I}_{TIE}(\xi,\eta)[1 + \gamma\pi\lambda R(\xi^2 + \eta^2)] - \qquad (7)$$
$$2(I_{TIE}\mu_{Born})^{\wedge}(\xi,\eta)\{\cos[\pi\lambda R(\xi^2 + \eta^2)] + \gamma\sin[\pi\lambda R(\xi^2 + \eta^2)]\}.$$

Noting that since $I_{TIE}(x,y)$ is by definition a slowly varying function, it is also possible to approximate $\hat{I}_{TIE}(\xi,\eta) \cong \hat{I}_{TIE}(\xi,\eta)\cos[\pi\lambda R(\xi^2 + \eta^2)]$ and $\hat{I}_{TIE}(\xi,\eta)\gamma\pi\lambda R(\xi^2 + \eta^2) \cong \hat{I}_{TIE}(\xi,\eta)\gamma\sin[\pi\lambda R(\xi^2 + \eta^2)]$. With the help of these further approximations, eq.(7) can be re-written as the linearized expression (Gureyev et al 2006)

$$\hat{I}_R(\xi,\eta) = \hat{I}_0(\xi,\eta)\{\cos[\pi\lambda R(\xi^2 + \eta^2)] + \gamma\sin[\pi\lambda R(\xi^2 + \eta^2)]\}. \qquad (8)$$

In view of this equation, the expression $\hat{\mathbf{G}}_{\gamma,R}(\xi,\eta) \equiv \cos[\pi\lambda R(\xi^2 + \eta^2)] + \gamma\sin[\pi\lambda R(\xi^2 + \eta^2)]$ can be regarded as a Fresnel "intensity propagator" (Nesterets et al 2016) in the case of homogeneous objects. The "phase" retrieval method based on eq.(8), i.e. $\hat{I}_0(\xi,\eta) = [\hat{\mathbf{G}}_{\gamma,R}(\xi,\eta)]^{-1}\hat{I}_R(\xi,\eta)$, represents a natural extension of the "homogeneous TIE" method (also known as the Paganin method (Paganin et al 2002)), which incorporates higher diffraction orders (Gureyev et al 2006). Note however that the Fresnel intensity propagator $\hat{\mathbf{G}}_{\gamma,R}(\xi,\eta)$ is still "incomplete", in the sense that it represents a first-order in $\gamma$ approximation to the full Fresnel propagator in the case of homogeneous objects.

Expanding the sine and cosine in eq.(7) into a Taylor series and taking into account that $\hat{I}_{TIE}(\xi,\eta) - 2(I_{TIE}\mu_{Born})^{\wedge}(\xi,\eta) \cong [I_{TIE}I_{Born}]^{\wedge}(\xi,\eta) = \hat{I}_0(\xi,\eta)$, we obtain:

$$\hat{I}_R(\xi,\eta) = \hat{I}_0(\xi,\eta)[1 + \gamma\pi\lambda R(\xi^2 + \eta^2)] - \qquad (9)$$
$$2[I_{TIE}\mu_{Born}]^{\wedge}(\xi,\eta)\sum_{m=1}^{\infty}\frac{(-1)^m[\pi\lambda R(\xi^2 + \eta^2)]^{2m}}{(2m)!}\left[1 + \frac{\gamma\pi\lambda R(\xi^2 + \eta^2)}{2m+1}\right].$$

This equation shows that the expression $\hat{I}_R^{(1)}(\xi,\eta) \equiv \hat{I}_R(\xi,\eta) - [1 + \gamma\pi\lambda R(\xi^2 + \eta^2)]\hat{I}_0(\xi,\eta)$ corresponds to the higher-order diffraction terms, which constitute the "non-TIE part" of the dark-field signal that we would like to find.

Note that different powers of the reciprocal-space variables $\xi$ and $\eta$ on the right-hand side of eq.(9) indeed correspond to distinct "diffraction orders", even though it may not be immediately obvious. Conventionally, diffraction orders with integer indexes ($m, n$) are associated with the Fourier components, $U_0^{(m,n)}(x,y) = \exp[i2\pi(mx/A + ny/B)]$, of the transmitted complex amplitude in the object plane, where the amplitude is assumed to be identically zero outside the rectangle $(-A/2 \leq x \leq A/2) \times (-B/2 \leq y \leq B/2)$. The Fourier transform of $U_0^{(m,n)}(x,y)$, i.e. $\hat{U}_0^{(m,n)}(\xi,\eta) = \delta(\xi - m/A)\delta(\eta - n/B)$, is non-zero only at the point $(\xi,\eta) = (m/A, n/B)$. It is easy to see, however, that if the Fourier spectrum $[I_{TIE}\mu_{Born}]^{\wedge}(\xi,\eta)$ behaves similarly to a Gaussian distribution $\exp[-(\xi^2 + \eta^2)/(2\sigma^2)]$, the terms $\hat{I}_0^{(m)}(\xi,\eta) \equiv (\xi^2 + \eta^2)^{2m}[I_{TIE}\mu_{Born}]^{\wedge}(\xi,\eta)$, with different indices $m = 1, 2, ...$, reach their maxima at the circles $(\xi^2 + \eta^2)/(4\sigma^2) = m$. This can be verified by noting that the function $(\xi^2 + \eta^2)^{2m}\exp[-(\xi^2 + \eta^2)/(2\sigma^2)]$ is equal to zero both at the origin of coordinates and at infinity, and the partial derivatives $\partial_\xi \hat{I}_0^{(m)}(\xi,\eta) = [4m(\xi^2+\eta^2)^{-1} - \sigma^{-2}]\xi \hat{I}_0^{(m)}(\xi,\eta)$ and $\partial_\eta \hat{I}_0^{(m)}(\xi,\eta) = [4m(\xi^2+\eta^2)^{-1} - \sigma^{-2}]\eta \hat{I}_0^{(m)}(\xi,\eta)$ are both equal to zero when $\xi^2 + \eta^2 = 4m\sigma^2$.

In order to show how the dark-field signal can be found in principle from the measured image intensity according to eq.(9), we substitute $\hat{I}_0(\xi,\eta) \cong [\hat{\mathbf{G}}_{\gamma,R}(\xi,\eta)]^{-1}\hat{I}_R(\xi,\eta)$ from eq.(8) into eq.(9) and leave only the lowest diffraction order of the term $I_{TIE}\mu_{Born}$. This produces the following approximate expression governing the fourth-order dark-field signal:

$$[\pi\lambda R(\xi^2+\eta^2)]^2[I_{TIE}\mu_{Born}]\hat{}(\xi,\eta) \cong \hat{I}_R(\xi,\eta) - [1 + \gamma\pi\lambda R(\xi^2+\eta^2)][\hat{\mathbf{G}}_{\gamma,R}(\xi,\eta)]^{-1}\hat{I}_R(\xi,\eta). \quad (10)$$

Taking the inverse Fourier transform of eq.(10), we can also obtain:

$$\mu_{Born}(x,y) \cong (2k/R)^2 \frac{\nabla_\perp^{-4}\{I_R(x,y) - [1-(\gamma/2)(R/k)\nabla_\perp^2](\mathbf{G}_{\gamma,R}^{-1}I_R)(x,y)\}}{I_{TIE}(x,y)}, \quad (11)$$

where $(\mathbf{G}_{\gamma,R}^{-1}f)(x,y) \equiv \iint \exp[i2\pi(x\xi+y\eta)]\hat{f}(\xi,\eta)\{\cos[\pi\lambda R(\xi^2+\eta^2)] + \gamma\sin[\pi\lambda R(\xi^2+\eta^2)]\}^{-1}dxdy$ is the real-space representation of the inverse Fresnel intensity propagator. Equation (11) corresponds to the lowest-order dark-field signal beyond the TIE-Hom approximation. Since all quantities on the right-hand side of eq.(11) can be calculated from the single measured image intensity $I_R(x,y)$, this equation can be solved for $\mu_{Born}(x,y)$, with the latter (small, but rapidly varying) term effectively representing the "source" of the dark-field signal in propagation-based imaging of homogeneous objects. As explained above, immediately after eq.(4), the same term can be also retrieved, alongside the phase term $\varphi_{Born}(x,y)$, in the case of general (non-homogeneous) objects. For that purpose, however, one would need to measure at least two image intensity distributions at two different object-to-image distances (Gureyev *et al* 2004).

Note also that, as can be seen from eq.(8), only even-order diffraction terms contribute to the Fresnel diffraction intensity under the considered conditions, while the odd-order terms cancel out due to the symmetry of the corresponding equations. The zero and the second diffraction orders are included in the TIE-Hom approximation, as shown in eq.(9). The lowest remaining diffraction order is the fourth one, and it can be extracted using eq.(11).

More generally, all diffraction orders beyond those included in the TIE approximation can be obtained by solving eqs.(4)-(6) for $I_{TIE}\mu_{Born}$ as explained above in the paragraph following eq.(6) (see also (Gureyev et al 2004)). The latter approach is employed in the next section for extracting the dark-field signal from experimental X-ray PBI CT images.

## 3. Experimental demonstration

Phase-contrast breast cancer imaging is an emerging medical imaging technology that is currently being developed and tested in several countries (Taba *et al* 2018). Because it requires intense spatially-coherent X-rays in order to be used effectively with humans, the method is predominantly implemented with synchrotron radiation (Longo *et al* 2016, Gureyev *et al* 2019), although experimental systems with micro-focus laboratory X-ray sources are also being tested (Havariyoun *et al* 2019). The dark-field modality of the GB method has been reported to be effective for analysis of microcalcifications (Rauch *et al* 2020). Here we demonstrate the improved visualization of microcalcification clusters in PBI CT breast cancer imaging performed at the Imaging and Medical Beamline (IMBL) of the Australian Synchrotron. The key experimental details are listed below and additional details can be found e.g. in (Gureyev *et al* 2019). The imaging experiment was conducted under a Human Ethics Certificate of Approval from Monash University and with written consent from the patient to image their clinical specimens.

A breast mastectomy sample was obtained from a breast cancer surgery operation and was imaged on the same day in a complete, intact, unfixed state at IMBL using the PBI CT technique. The professional pathology analysis of this specimen later reported a high-grade ductal carcinoma in-situ and a grade 3 invasive carcinoma of non-specific type in this specimen. The report also listed scar tissue formation with calcifications. The specimen contained several surgical clips that were clearly visible in the CT images. The primary question that we intended to address by the present study was how to improve the assessment of micro-calcifications, which are not always easy to resolve and evaluate in detail in conventional absorption CT or bright-field PBI CT images.

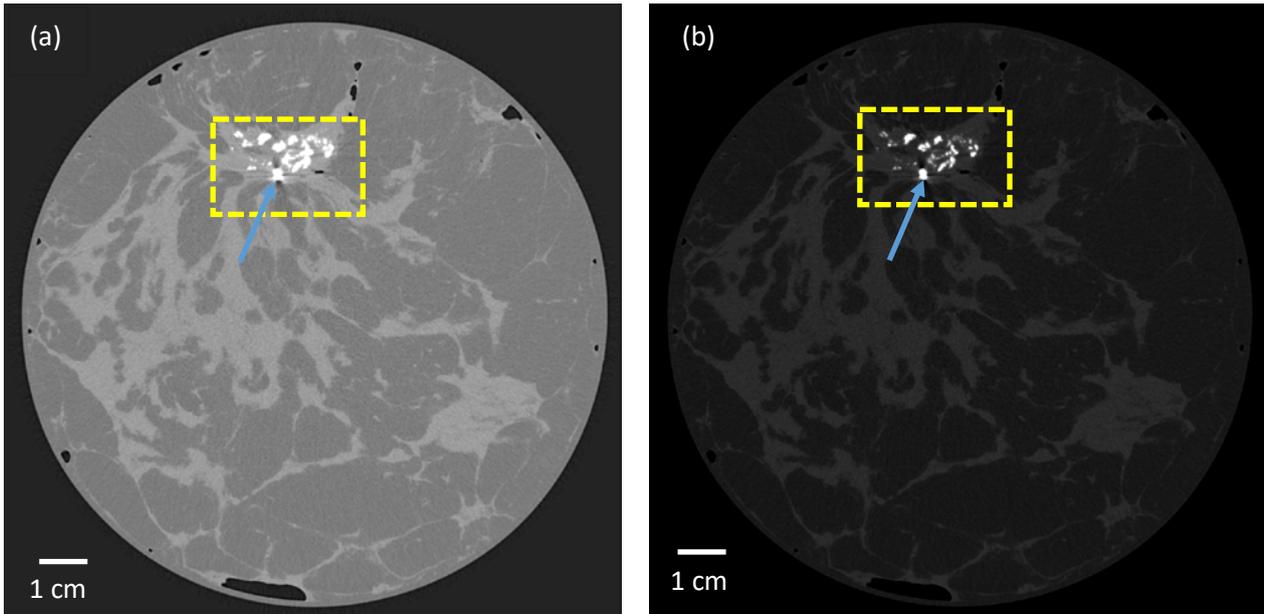

Figure 1. PBI CT reconstruction of mastectomy sample 4607971L from X-ray projections collected at 34keV energy, 6 m sample-to-detector distance and 24mGy dose. A coronal slice from the reconstructed bright-field CT image stack: (a) default image histogram setting; (b) image histogram setting adjusted to optimally visualize the microcalcifications. The dashed-line box indicates the area occupied by a microcalcification cluster. The blue arrow points to a surgical clip.

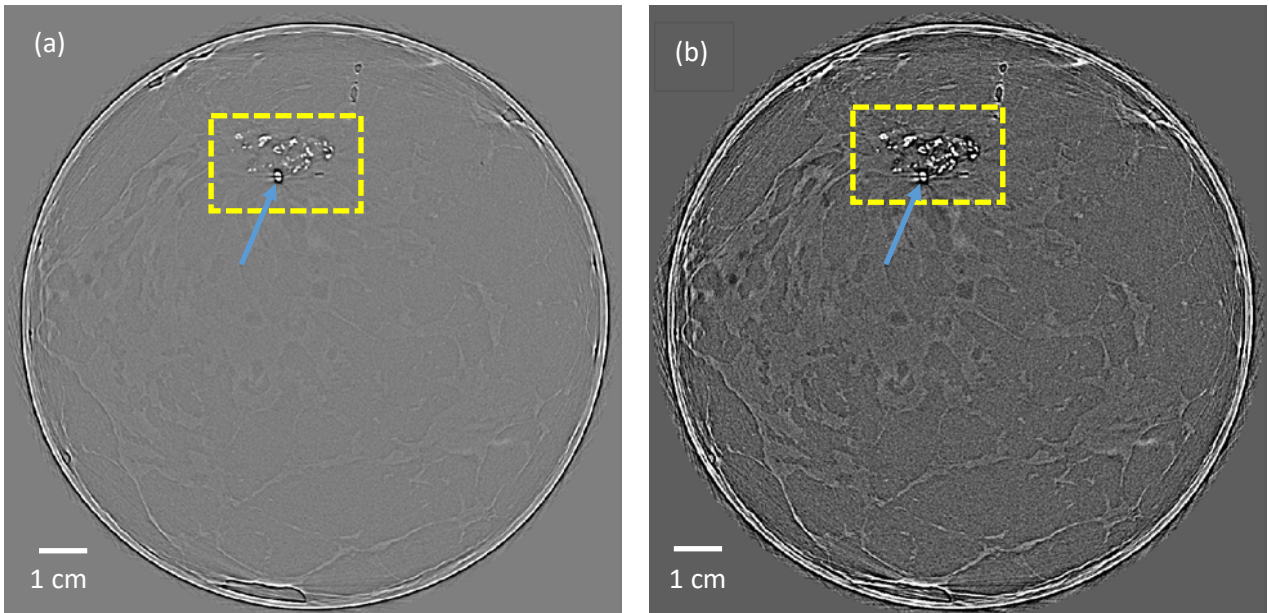

Figure 2. A slice reconstructed at the same location from the same PBI CT dataset as in Fig.1, but using the proposed dark-field reconstruction method: (a) default image histogram setting; (b) image histogram setting adjusted to optimally visualize the microcalcifications. The dashed-line box and the arrow have the same meaning as in Fig.1.

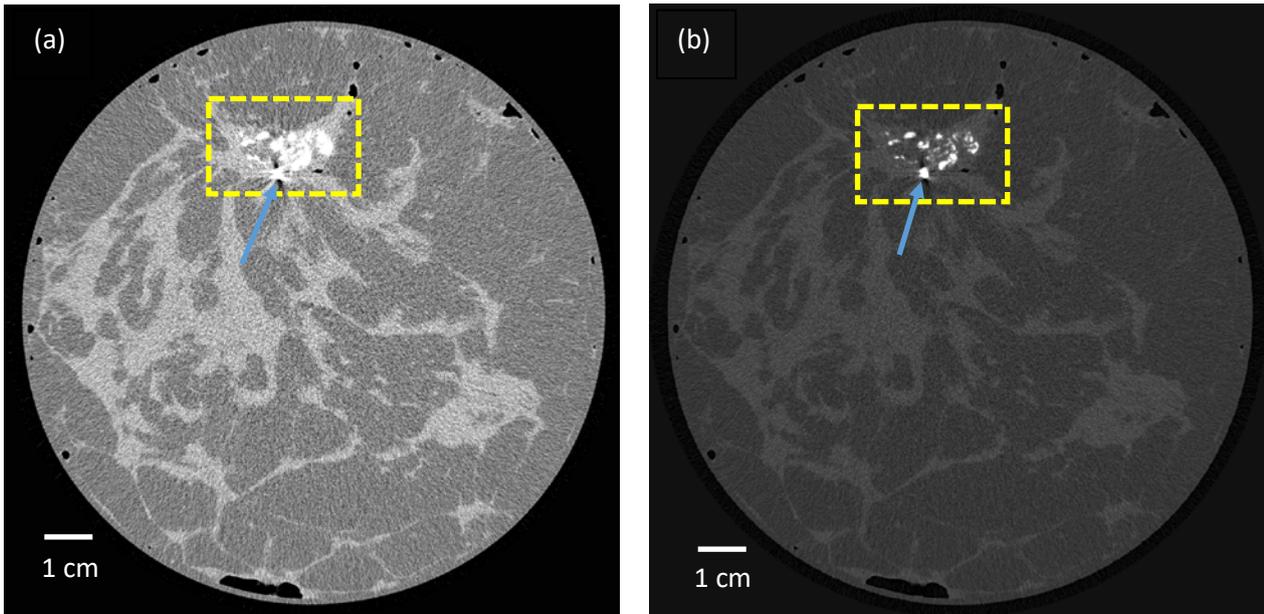

Figure 3. PBI CT reconstruction of mastectomy sample 4607971L from X-ray projections collected at 34keV energy, 6 m sample-to-detector distance and 4mGy dose. A coronal slice (at the same location as in Fig.1) from the reconstructed bright-field CT image stack: (a) default image histogram setting; (b) image histogram setting adjusted to optimally visualize the microcalcifications. The dashed-line box and the arrow have the same meaning as in Fig.1.

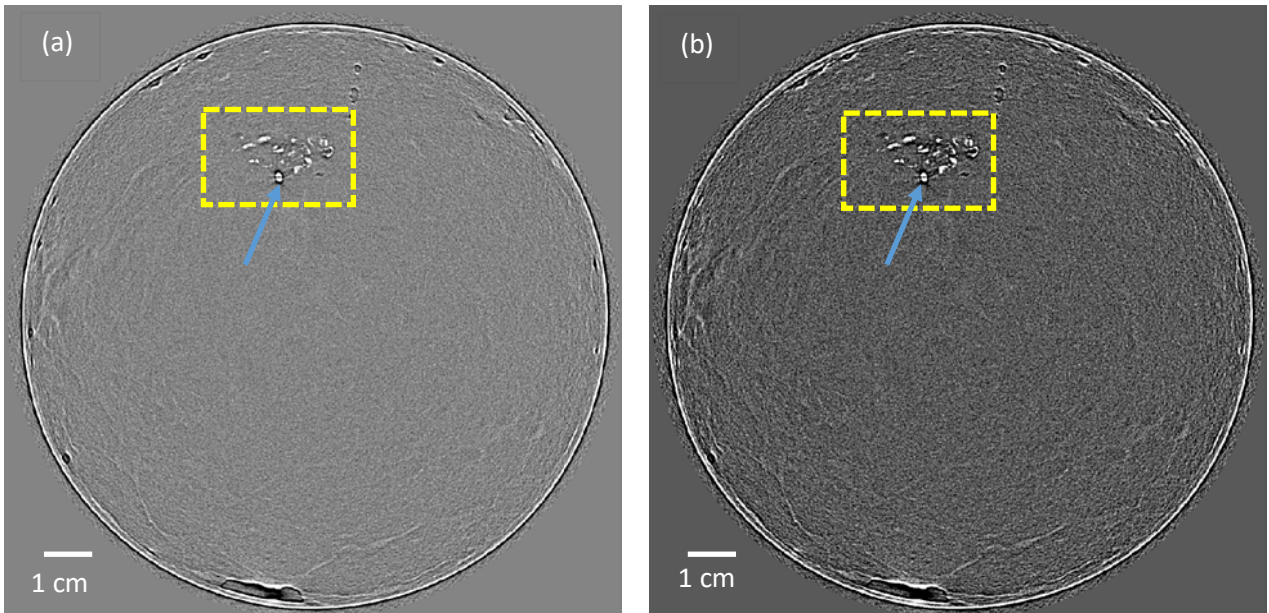

Figure 4. A slice reconstructed at the same location from the same PBI CT dataset as in Fig.3, but using the proposed dark-field reconstruction method: (a) default image histogram setting; (b) image histogram setting adjusted to optimally visualize the microcalcifications. The dashed-line box and the arrow have the same meaning as in Fig.1.

The two PBI CT scans analyzed here were collected at 24 mGy mean glandular dose distributed evenly between 4,800 projections over 180 degrees, and at 4 mGy / 2,400 projections, respectively. Both scans were carried out using quasi-plane monochromatic X-rays with energy E = 34 keV and a free-space propagation distance of 6 m between the sample and the detector (which corresponded to the effective defocus distance of 5.748 m when the relevant geometric magnification of the imaging setup was taken into account). The X-ray detector used for these scans was a Hamamatsu CMOS Flat Panel Sensor C10900D with pixel size 100 μm × 100 μm and field of view 12.16 cm (horizontal) × 12.32 cm (vertical). The spatial resolution of approximately 150 μm (horizontal) × 170 μm (vertical) was determined predominantly by the detector's point-spread function, while the slight asymmetry of the resolution was due to the anisotropic X-ray source at IMBL.

Figure 1 presents the bright-field PBI CT image of a single coronal slice of the mastectomy sample reconstructed from a scan with 24 mGy dose using the TIE-Hom method (Gureyev et al 2019). While the microcalcifications are not well visualized in the default image setting in Fig.1(a), the same microcalcification cluster is comparatively well resolved in Fig.1(b) where the histogram has been specifically adjusted to maximize the visibility of these microcalcifications. In a real-life radiological examination setting such an adjustment may require the observer to successfully detect the microcalcifications in the default setting in the first place (where the microcalcifications may be partially masked by the dense glandular or cancerous tissue and by the CT artefacts produced by the adjacent surgical clips), and then adjust the image display parameters accordingly. Such requirements may lead to missed features and to extra time spent on the examination. In comparison, Fig.2 depicts the same slice reconstructed from the same CT projection data as in Figure 1, but using the dark-field mode algorithm as defined by eqs.(5)-(6) above. The visibility of the microcalcification cluster is already quite good in the default image setting here (Fig.2(a)), and the visibility becomes even better after the appropriate histogram adjustment. The shape of individual microcalcifications is resolved better in Fig.2(b), compared to Fig.1(b), which is important for breast cancer diagnosis (, Nalawade 2009, Rauch et al 2020). Figures 3 and 4 present the same PBI CT slice of the same sample as in Figs.1 and 2, but here the CT images have been obtained from a PBI CT scan with a more clinically relevant X-ray dose of 4 mGy per CT scan. Despite the six-fold reduction in the dose, the microcalcifications remain highly visible in these images, even though the image noise is obviously higher in Figs.3-4, compared Figs.1-2. The conclusion about the easier detectability and better edge definition of the microcalcifications in the dark-field PBI CT mode, compared to the bright-field mode, is still clearly valid at this lower dose. Note, in particular, that the two "legs" of the surgical clip indicated by a blue arrow in each figure are not resolved in the bright-field images, but they are clearly resolved in the dark-field images, even at the lower dose (Fig.4(a) and (b)).

## 4. Discussion

We have proposed a method for dark-field signal extraction from 2D and 3D PBI images, which does not require any additional image acquisitions compared to the usual bright-field imaging implementation of this method. Instead, our method employs a specialized algorithm for image processing that allows one to extract high-spatial-order diffraction components from the bright-field images. In particular, in the case of samples that can be considered approximately homogeneous for the purpose of PBI image analysis, the new method requires only a single image in 2D setting and only one image per projection angle in the 3D PBI CT setting. This may represent an important practical advantage compared to dark-field imaging modalities previously described in the context of other X-ray phase-contrast methods, such as GB, EI or ABI, where the acquisition of dark-field images inevitably required different positioning of the "wavefront analyser" elements compared to bright-field imaging. Our method extracts both the bright-field and the dark-field images of the sample from a single experimental image collected in conventional PBI setting. The dark-field images tend to show more of the SAXS signal, compared to bright-field images, when a non-negligible amount of SAXS signal is actually produced by a given sample. The fact that proposed dark-field PBI method visualizes SAXS signals more clearly, compared to the bright-field PBI, can be beneficial for medical imaging of lungs and other vital organs where the increased scattering may point to areas of abnormal tissue.

As a first example of a potential practical application, we have demonstrated that, in the case of human breast cancer imaging, the proposed dark-field method is capable of delivering better-quality images showing the shape and quantity of microcalcifications which are often used to differentiate between benign and malignant tissues in breast cancer diagnosis. Whilst one may argue that with conventional windowing solutions available on any CT console, visualisation of the microcalcifications is possible, this relies on two assumptions. Firstly that the operator or clinician has the skills to use the appropriate windowing level, and secondly that there is enough time to make these adjustments. With regard to the latter, clinicians are under immense pressure to read very large amounts of images, with 400 slices not being uncommon for one CT run, therefore having the luxury of windowing appropriate slices in a comprehensive way may not be possible. Instead clinicians need rapid presentations of important pathologies and having a dark-field mode available with a single button highlighting important calcification clusters presents a reasonable and potentially very useful solution. This dark-field facility would also encourage more rapid identification of the microcalcification features that describe malignant lesions such as clustering, pleomorphism and rod shapes. The method has the potential to improve imaging of breast cancers in patients with breast tissue that presents challenges for conventional imaging, such as recurrent cancers along scar lines, and cancers in tissue treated with radiation therapy. The usefulness of this alternative approach for clinicians will need to be confirmed to make sure that opting for a micro-calcification visualiser does not in any way have an inadvertent affect on normal expert behaviour and recognition of other important features. Our subsequent work will involve a systematic analysis of the radiological value of the dark-field method in breast PBI CT imaging, in particular. Applications to PBI imaging of the lungs are likely to be of interest as well, as indicated by recent publications (Wagner *et al* 2018, Fingerle *et al* 2019).

## Acknowledgements

The experimental part of this research was undertaken on the Imaging and Medical beamline at the Australian Synchrotron, part of ANSTO. The following funding is acknowledged: Australian National Breast Cancer Foundation (grant No. IN-16-001), and Project Grant APP1138283 from the National Health and Medical Research Council (NHMRC), Australia. We acknowledge the support of Drs. J. Fox, B. Kumar and Z. Prodanovic of Monash University in providing the mastectomy samples for this study.